\documentstyle[12pt]{article}

\input{epsf}

\newcommand{\energy}[1]{F_{\rm #1}}

\begin{document}
\begin{titlepage}
\begin{center}

\today    \hfill       MIT-CTP-2297
\hskip .5truein       \hskip .5truein hep-th/9404175\\

\vskip .2in

{{\Large \bf Counting Strings and Phase Transitions in 2D QCD
\footnote{This work was supported in part by  funds provided by the
U.S. Department of Energy (D. O. E.)  under contract
DE-AC02-76ER03069, and in part by the divisions of Applied Mathematics of
the U.S. Department of Energy under contracts DE-FG02-88ER25065 and
DE-FG02-88ER25066.}}}

\vskip .15in

Washington Taylor\footnote{\tt wati@mit.edu}\\

{\small {\em
Center for Theoretical Physics \\
Laboratory for Nuclear Science \\
and Department of Physics\\
Massachusetts Institute of Technology \\
Cambridge, Massachusetts 02139, USA}}
\end{center}
\vskip .15in

\begin{abstract}
Several string theories related to QCD in two dimensions are studied.
For each of these theories the large $N$ free energy on a (target)
sphere of area $A$ is calculated.  By considering theories with
different subsets of the geometrical structures involved in the full
QCD${}_2$ string theory, the different contributions of these
structures to the string free energy are calculated using both
analytic and numerical methods.  The equivalence between the leading
terms in the $SU(N)$ and $U(N)$ free energies is simply demonstrated
from the string formulation.  It is shown that when $\Omega$-points
are removed from the theory, the free energy is convergent for small
and large values of $A$ but divergent in an intermediate range.
Numerical results indicate that the free energy for the full QCD${}_2$
string fails to converge at the Douglas-Kazakov phase transition
point.  Similar results for a single chiral sector of the theory, such
as has recently been studied by Cordes, Moore, and Ramgoolam, indicate
that there are three distinct phases in that theory.  These results
indicate that from the point of view of the strong coupling phase, the
phase transition in the full QCD${}_2$ string arises from the entropy
of branch-point singularities.
\end{abstract}
\end{titlepage}
\newpage
\renewcommand{\thepage}{\arabic{page}}
\setcounter{page}{1}

\section {Introduction}
\setcounter{equation}{0}
\baselineskip 18.5pt

In the last year, there has been a resurgence of interest in 2D gauge
theories.  The Yang-Mills theories on an arbitrary compact 2-manifold
${\cal M}$ of genus $G$ and dimensionless area $A$, with gauge groups
$SU(N)$, $U(N)$, $Sp(N)$, and $SO(N)$, have been shown to have
partition functions with asymptotic expansions in $1/N$ which are
precisely equal to the partition functions of simple string theories
[1-6].  These string partition functions are expressed as sums over
covering maps
\begin{equation}
\nu: {\cal N} \rightarrow {\cal M},
\end{equation}
from 2-manifolds ${\cal N}$ of arbitrary genus $g$ onto ${\cal M}$,
where the covering maps have a finite number of singular points of
various types whose geometry depends upon the choice of genus $G$ and
the gauge group.  An expectation value of a Wilson loop in such a 2D
gauge theory can also be expressed in terms of the string picture as a
sum over open string maps where the boundary of the string world sheet
is mapped to the Wilson loop.  For a review of this work, see
\cite{gt3}.

One of the primary goals of working on a string description for 2D QCD
is to find a formulation of the theory which can be extended to higher
dimensions, so that we might develop a string picture of 4D QCD which
would allow us to calculate nonperturbative results in the physical
theory.  In order to extrapolate the QCD${}_2$ string theory to higher
dimensions, it seems that it will be necessary to develop a world
sheet action formalism for this theory; recently, a great deal of
progress has been made in this direction by Cordes, Moore, and
Ramgoolam \cite{cmr} (henceforth CMR), and by Horava \cite{horava}.

Even if a good string description can be found for the asymptotic
$1/N$ expansion of the 4D theory, however, it is still a nontrivial
task to relate this asymptotic expansion to the theory for finite
values of $N$.  One example of a difficulty which may arise in finding
such a relation is the possible existence of large $N$ phase
transitions.  It has long been known that in the lattice formulation,
the large $N$ description of QCD can suffer a phase transition between
the strong and weak coupling regimes \cite{gw}.  One possible way to
avoid this difficulty is to directly formulate a string theory for the
finite $N$ gauge theory, as was done in \cite{bt} for the QCD${}_2$
string.  However, it seems unlikely that this approach will lead to a
reasonable string picture in four dimensions.  Thus, it is important
to understand in what cases phase transitions arise in the two
dimensional theory, and what the implications of such phase
transitions are for the physical four dimensional theory.

Last year, Douglas and Kazakov (DK) addressed this problem by
considering the leading term $F (A) N^2$ in the free energy of the
$U(N)$ gauge theory on the two dimensional sphere \cite{dk}.  Whereas
the leading terms of the free energy on higher genus surfaces are
easily seen to be well-behaved, an infinite number of terms can
contribute to the quantity $F (A)$, and there is the potential for a
large $N$ phase transition to arise.  Indeed, DK were able to show, by
finding the saddle point of the matrix model corresponding to the
exact expression for the QCD${}_2$ partition function, that at the
point $A =\pi^2$ there is a phase transition.  In the large area phase
$F (A)$ is correctly described by the string expansion, whereas in the
small area phase this free energy becomes trivial.  It was
subsequently pointed out by Minahan and Polychronakos
\cite{mp} that the analysis of DK relied upon an Ansatz which
essentially restricted the theory to the sector with $U(1)$ charge $Q
= 0$.  For a full treatment of the free energy in the $U(N)$ theory,
they showed that it is necessary to sum over a range of values of
$Q$, for each of which there is a distinct saddle point.  More
recently, the DK phase transition was studied from the point of view
of the weak coupling phase by Gross and Matytsin, and it was shown
that the phase transition is induced by instantons \cite{gm}.  In that
work it was also argued that in four dimensions no analogous phase
transition should occur.  The  description of the 2D Yang-Mills
partition function in terms of instantons was originally discussed by
Witten \cite{witten}, and the role of these instantons in the DK phase
transition was  suggested in \cite{cdmp}.

The work presented in this paper is in a sense complementary to that
of \cite{gm}, in that here the QCD${}_2$ theory is studied from the
point of view of the strong coupling phase.  The goal is to attain a
deeper understanding of how the large $N$ phase transition in $F (A)$
arises from the string point of view.
As a step towards this goal, we wish to ascertain
which structures in the string theory are responsible for the finite
radius of convergence of the string expansion.  We
consider a general class of string theories of the type of the
QCD${}_2$ string.  In the string theories associated with $U(N)$ and
$SU(N)$ gauge theories, there are several types of singularities
allowed in the string maps.  The most basic singularities are simple
branch-points which connect a pair of sheets of the covering space at
a point in the target.  In addition, the $SU(N)$ theory allows
infinitesimal tubes which connect pairs of sheets which can be either
of identical or opposite orientation, and also entire world sheet
handles which are mapped to single points in the target space.
Finally, when $G \neq 1$, there are additional singularities allowed
at special twist points called $\Omega$-points and
$\Omega^{-1}$-points; these last singularities seemed originally
somewhat complicated from the string perspective, but have recently
been understood as simply arising from orbifold Euler characters of
the relevant spaces of branched covering maps \cite{cmr}.  In this
work, we consider string theories in which different subsets of these
singularity types are allowed; we also consider some theories in which
string maps are constrained to be orientation-preserving (this
corresponds to taking a single ``chiral'' sector of the full theory in
which maps can be locally orientation-preserving or -reversing).

The results of the investigations carried out in this paper are that
the convergence of the string expansion depends strongly upon what
singularity types are allowed in the string theory, and that the radii
of convergence of these expansions are apparently related to phase
transition points in the associated theories.  When branch-points are
neglected, and all singularities are taken to arise from
$\Omega$-points, the string expansion of the free energy $F (A)$
converges to a smooth function for all $A > 0$ and hence the theory
has no phase transitions; this theory is closely related to the
topological string which formed a starting point in \cite{cmr}.  If we
neglect $\Omega$-points, on the other hand, and only include
branch-point type singularities in the string theory, we find that the
free energy converges for small and large values of $A$, but diverges
in an intermediate region.  We also consider the chiral theory
containing both $\Omega$-points and branch-point singularities.
Although it is difficult to describe this theory analytically, we
carry out partial summations of the string expansion and graph the
results.  We find that as in the theory with only branch points, in
the regions of small and large area the free energy is convergent,
while in an intermediate region the expansion fails to converge.
Similar numerical analysis of the full QCD${}_2$ string theory shows a
convergence of the string expansion above the DK critical point $ A
=\pi^2$ and a divergence below this point.  This result suggests that
the strong coupling string expansion contains nonperturbative
information which describes the phase transition point of the theory.
If the chiral string theory behaves similarly, we would expect to see
3 distinct phases in that theory.  The behavior of the theory without
$\Omega$-points indicates that it is the entropy factor associated
with branch-point singularities which is ultimately responsible for
the failure of the string expansion to converge at the phase
transition point.

Another result which follows naturally from the string picture is the
observation that when both orientation-preserving and
orientation-reversing tubes are included in the string theory, the
contributions from these objects to the leading term in the free
energy exactly cancel.  Thus, we have a simple geometrical
demonstration of the result that the free energy $F (A)$ in the
$SU(N)$ theory is precisely the same as the free energy in the $Q = 0$
sector of the $U(N)$ theory in the large area phase, and thus that the
DK phase transition describes correctly the complete $SU(N)$ theory.

In Section 2, we review briefly the rules for counting
covering maps in the QCD${}_2$ string theory, and we define a set of
simplified QCD${}_2$-type string theories.  In Section 3, several
combinatorial problems are discussed which are related to  calculating
the free energy of QCD${}_2$-type strings.  In Section 4 we perform
the central calculations in the paper, describing the free energy in
the string theories of interest using both analytic and numerical
techniques.  The convergence properties of these string sums are
described and the consequences for the phase structure of the theories
are discussed.  In this section we also prove the equivalence of the
$SU(N)$ and $U(N)$ free energies. Finally, in Section 5 we conclude
with a discussion of the implications of this work and related open
questions.

\section{QCD${}_2$ Strings}
\setcounter{equation}{0}
\baselineskip 18.5pt

We review here the essential features of the sum over covering maps
describing the QCD${}_2$ string theory.  The results, which are
stated here without proof, are proven in
\cite{gt1,gt2}.

The partition function for the 2D Yang-Mills theory on a manifold
${\cal M}$ of genus $G$ and area $A$ is given by
\cite{migdal,rusakov}
\begin{eqnarray}
Z(G, A,N) & = &  \int[{\cal D} A^\mu]
{\rm e}^{- {1\over 4 g^2} \int_{\cal M}
 {\rm Tr} (F \wedge \star F)} \nonumber\\
  & = & \sum_{R} (\dim R)^{2 - 2G}
  {\rm e}^{-\frac{A}{2 N}C_2(R)}, \label{eq:partition}
\end{eqnarray}
where the sum is taken over all irreducible representations of the
gauge group, with $\dim R$ and $C_2(R)$ being the dimension and
quadratic Casimir of the representation $R$.  We absorb the coupling
constant $\lambda = g^2 N$ into the dimensionless area $A$ for
notational convenience.  The partition function can be rewritten in
the form
\begin{equation}
Z (G, A, N) = \int_{\Sigma({\cal M})} \,{\rm d} \nu\, W (\nu),
\end{equation}
where $\Sigma ({\cal M})$ is a set of branched covering maps of ${\cal
M}$ by oriented 2-manifolds ${\cal N}$ of arbitrary genus $g$.  The
weight of each map in the partition function is given by
\begin{equation}
W (\nu) =   \pm
\frac{N^{2-2 g}}{ | S_\nu |}
{\rm e}^{- \frac{n  A}{2}},
\end{equation}
where $n$ is the degree of the map $\nu$ and $| S_\nu |$ is the
symmetry factor of the map (the number of diffeomorphisms $\pi$ of
${\cal N}$ which satisfy $\nu \pi = \nu$).  The sign of the weight
depends upon the set of singular points in the map $\nu$.  Each map
$\nu$ describes a covering of ${\cal M}$ by a number of
orientation-preserving sheets and a number of orientation-reversing
sheets.  We refer to these two types of sheets as belonging to two
chiral sectors of the theory.

\begin{figure}
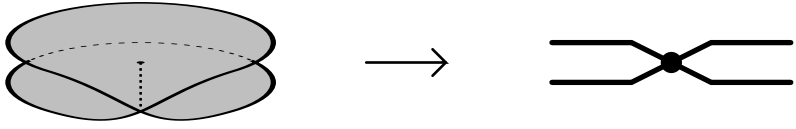

\vspace{84pt}

\caption[x]{\footnotesize Schematic representation of a branch-point}
\label{f:branch}
\end{figure}

The maps in $\Sigma ({\cal M})$ may
include the following  types of singular points:

\vspace{0.2in}

$\bullet$ {\bf simple branch-points}: These points are topologically
equivalent to the branch-points occurring in the map $z \rightarrow
z^2$ from the Riemann sphere $S^2$ to itself.  Given a closed path
$\gamma$ around a simple branch-point on the target space, and a
labeling of the sheets of the cover at a point $p \in \gamma$, the
lift of $\gamma$ to the covering space (string world sheet) induces a
permutation on the sheet labels which contains a single cycle of
length 2 and $n-2$ cycles of length 1.  Branch-points carry a factor
of $-1$ in the string map weight.  Furthermore, a branch-point may
occur anywhere on the target space, so the positions of these
singularities form a set of modular parameters for branched covers
which must be integrated with a measure locally proportional to the
area measure on the target manifold.  A schematic representation for a
simple branch-point is shown in Figure~\ref{f:branch}.

\begin{figure}
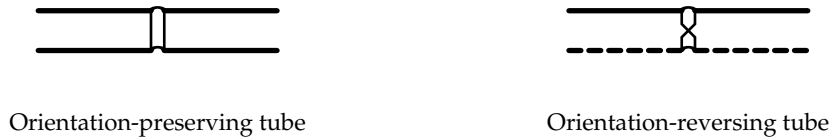

\vspace{85pt}

\caption[x]{\footnotesize Schematic representation of infinitesimal tubes}
\label{f:tubes}
\end{figure}

$\bullet$ {\bf Infinitesimal tubes:} An infinitesimal tube is a
singularity which occurs when a nontrivial loop in the string world
sheet is mapped to a single point in the target space.  Such a tube
can  connect two sheets of either the same or opposite orientation, in
which case it is referred to as an orientation-preserving or
-reversing tube respectively.  The lift of a target space curve around
a tube gives the identity permutation on the sheets of a cover.
Infinitesimal tubes can occur at any point on the target manifold;
thus, the positions of tubes are an additional set of modular
parameters for the space of covers, again carrying a measure locally
proportional to the target space area measure.  Furthermore,
orientation-reversing tubes carry a factor of $-1$.  A schematic
representation of both types of tubes is given in
figure~\ref{f:tubes}; in this figure, as in the remainder of this
paper, we denote orientation-preserving sheets of a covering space
with a solid line, and orientation-reversing sheets with a dashed
line.

$\bullet$ {\bf Infinitesimal handles:} Just as infinitesimal tubes are
singularities where a circle in the world sheet goes to a point in the
target space, an infinitesimal handle occurs when an entire handle of
a string world sheet is mapped to a single point in the target space.
Again, the positions of such singularities must be integrated over the
target space.  Because in this paper we are essentially only concerned
with maps from $S^2\rightarrow S^2$, this type of singularity will not
play a role in the calculations performed here.

\begin{figure}
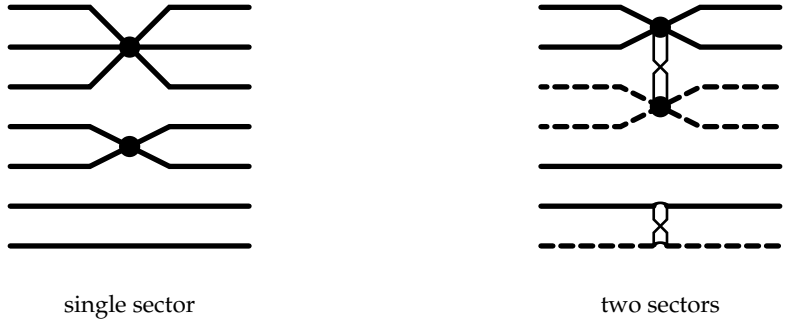

\vspace{155pt}

\caption[x]{\footnotesize $\Omega$-point singularities in 1 and 2 sectors}
\label{f:op}
\end{figure}

$\bullet$ {\bf $\Omega$-points:} An additional type of singularity can
occur when the target space is a sphere.  An $\Omega$-point is a {\it
fixed} point on the target space at which the string maps can have a
singularity structure combining multiple branch-point singularities
with orientation-reversing tubes.  In a single chiral sector of the
theory, the singularity at an $\Omega$-point is a multiple
branch-point singularity which is described by an arbitrary
permutation on the sheets of the covering space when following the
lift of a closed target space curve $\gamma$ around the
$\Omega$-point.  In the coupled theory, an $\Omega$-point allows the
same types of multiple branch-point singularities in each chiral
sector; in addition, each connected cycle of sheets may be connected
by an orientation-reversing tube to a cycle of the same length in the
opposite sector.  Each such orientation-reversing tube contributes a
factor of $-1$ to the weight of the map.  Because the $\Omega$-points
are fixed on the target space, their positions are not integrated over
and do not give rise to factors of the area.  The geometrical
structure of example $\Omega$-point singularities in 1 and 2 sectors
are shown schematically in figure~\ref{f:op}; note that we do not
bother to include in this schematic picture information about the
precise permutation at the $\Omega$-point, simply which sheets are
connected by cycles.  When the target space is a sphere, there are two
points with $\Omega$-point singularities.

$\bullet$ {\bf $\Omega^{-1}$-points:} When the genus of the target
space is greater than 1, fixed singularity structures similar to
$\Omega$-points occur, which are essentially inverses  of those
objects.  At an $\Omega^{-1}$-point singularity, there can be an
arbitrary number of $\Omega$-point type singularities contracted to a
single point, each carrying a factor of $-1$.  Because we will only be
interested in results on the sphere in this paper, we will not discuss
$\Omega^{-1}$-points further here.

\vspace{0.2in}

This completes the list of singularity types which may appear in the
string formulation of the gauge theories with gauge groups $U(N)$ and
$SU(N)$.  There are additional types of singularities which must be
considered for the gauge groups $SO(N)$ and $Sp(N)$ \cite{nrs,r}; in
addition, a formulation of the finite $N$ gauge theory in terms of
strings necessitates the introduction of a ``projection'' point
singularity, similar to the $\Omega$-point singularity \cite{bt}.

Although we only have a gauge theoretic description of the string
theories with certain combinations of singular points allowed in the
space of maps $\Sigma({\cal M})$, it is possible to define a string
theory with an arbitrary combination of allowed singularity types.  We
will also find it interesting to discuss ``chiral'' string theories in
which the maps in $\Sigma$ are restricted to be
orientation-preserving.  Thus, we define a string partition function
$Z_{\alpha i} (G,A, N)$, where $\alpha \subseteq \{B,H,O,T,
\tilde{T}\}$, $i \in\{1,2\}$ to be a string theory where the indicated
set $\alpha$ of singularity types is allowed (branch-points, handles,
$\Omega$-points, orientation-preserving tubes, orientation-reversing
tubes), and with $i$ sectors.  For example, $Z_{BO1}$ is the string
partition function for a theory with a single sector and branch-point
and $\Omega$-point singularities but no handles or tubes.
The specific results for the gauge
groups $SU(N)$ and $U(N)$ are that
\begin{eqnarray}
Z_{SU(N)} & = &  Z_{BHOT \tilde{T}2} \nonumber\\
Z_{U(N)^{Q = 0}} & = &  Z_{BO2}.
\end{eqnarray}
The chiral $SU(N)$ and $U(N)$ theories $Z_{BHOT1}$ and $Z_{BO1}$ are
also of particular interest, as the large $N$ gauge theories almost
factorize into two copies of the chiral theories.  These theories are
used to construct the string formulation of finite $N$ theories
\cite{bt}, and also arise naturally in a topological field theory
approach to the QCD${}_2$ string \cite{cmr}.

\section{Combinatorial Problems}
\setcounter{equation}{0}
\baselineskip 18.5pt

In this section we will briefly describe two simple combinatorial
problems and their solutions, which will be of assistance in
calculating the string summations in the free energies of the theories
$Z_{\alpha i}$.  These problems can each be stated either in a
geometric or combinatorial fashion.

\vspace{0.2in}

\noindent {\bf Problem 1a:} {\it
What is the number $H_n$ of inequivalent holomorphic maps of degree
$n$ going from $S^2 \rightarrow S^2$ with $n-1$ simple branch-points
whose images are fixed points $z_1, \ldots,z_{n-1},$ and with a
branch-point of multiplicity $n-1$ whose image is at $\infty$?  }

(we consider two maps $\nu, \mu$
to be equivalent when there exists an $SL(2,C)$ transformation $\zeta$
with $\nu = \mu \zeta$.)

\vspace{0.2in}

The combinatorial version of this problem is

\vspace{0.2in}

\noindent {\bf Problem 1b:} {\it
What is the number $C_n$ of connected graphs which can be formed with
$n$ labeled vertices and $n-1$ edges?}

\vspace{0.2in}

That these are in fact equivalent questions can be seen through the
following construction.  Given a map which satisfies the conditions of
Problem 1a, we can introduce a labeling of the sheets of the cover at
a fixed (unbranched) point in the target.  From this point we can
choose a set of canonical paths in the target homotopic to loops
around the points $z_i$.  Each of these paths can then be associated
with a permutation of two labeled sheets, say $s$ and $t$, and thus
with an edge in a graph connecting vertices labeled $s$, $t$.  The
graph resulting from this construction must be connected since the
product of the permutations is a cyclic permutation on $n$ elements.
This gives us a map from the set of solutions to the first problem
with labeled sheets, to the set of solutions to the second problem
with labeled edges.  The construction can easily be inverted to show
that the map is 1-1, using the fact that every topological type of
branched cover of a Riemann surface with fixed branch-points has
precisely one holomorphic representative.  Since the number of
possible labels on sheets is $n!$ and the number of possible labelings
of edges is $ (n-1)!$, we find that $H_n = C_n/n$.

The solution to Problem 1b was found in 1889 by Cayley \cite{cayley}
in the course
of an investigation of hypothetical chemical structures, and is given
by
\begin{equation}
C_n = n^{n-2}.
\label{eq:solution1}
\end{equation}

\vspace{0.2in}

\noindent {\bf Problem 2a:} {\it
What is the number $G_n$ of inequivalent holomorphic maps of degree
$n$ going from $S^2 \rightarrow S^2$ with $2n-2$ simple branch-points
whose images are fixed points $z_1, \ldots,z_{2n-2}$?  }

\vspace{0.2in}

Note that when $n = 2$, the single holomorphic map contributing to
$G_n$ has a symmetry factor of 2, so that $G_2 = 1/2$; for all other
values of $n$, the symmetry factors of all maps are $1$.  The
combinatorial version of this problem is

\vspace{0.2in}

\noindent {\bf Problem 2b:} {\it
What is the number $P_n$ of sequences $a_1, \ldots,a_{2n-2}$ of
elementary transposition permutations in $S_n$ which satisfy $a_1
\cdots a_{2n-2}= 1$ and which generate the full permutation group?}

\vspace{0.2in}

The  solutions of these two versions of the problem are related by $
P_n = G_n n!$,  as can be shown by an argument similar to that
relating the two versions of Problem 1.

The answer to problem 2 is given by
\begin{equation}
P_n=n^{n-3} (2n-2)!.
\label{eq:solution2}
\end{equation}
This result, which to the best knowledge of the author, is not known
in the mathematical literature, was conjectured  on the basis of a
numerical calculation of $G_n$ for small values of $n$ by D. Gross and
the author.  The result can be proven using a matrix model for chiral
2D QCD, and will be presented in a separate publication \cite{ct}.

The essential conclusions of the present work do not in fact rely upon
the exact formula (\ref{eq:solution2}), but rather upon its asymptotic
properties as $n \rightarrow \infty$.  Thus, for the purposes of this
paper it will be sufficient to prove a simple set of bounds on $P_n$
which determine its asymptotic form.  We now show that
\begin{equation}
n^{2n-4} (n-1)!  \leq P_n \leq n^{2n-4} (2n-2)!/(n-1)!.
\label{eq:inequalities}
\end{equation}
These inequalities follow using the result (\ref{eq:solution1}), which
is equivalent to the statement that the number of sequences $a_1,
\ldots,a_{n-1}$ of elementary transpositions whose product $a_1 \cdots
a_{n-1}$ is a {\em given} cyclic permutation on $n$ elements is
precisely $n^{n-2}$.  We can now enumerate the subset of solutions of
Problem 2b which have the property that the first $n-1$ of the $2n-2$
permutations generate the group.  Any set of $n-1$ elementary
transpositions which generate the group have a product which is a
cyclic permutation on $n$ elements; thus, we can characterize each
solution in the subset by a cyclic permutation $\pi$, of which there
are $(n-1)!$, and two sets of $n-1$ permutations ($a_1,
\ldots,a_{n-1}$ and $a_{n}, \ldots a_{2n-2}$) each of whose product is $\pi$.
There are thus $n^{2n-4}(n-1)!$ solutions which satisfy this
condition, so we have the first inequality in (\ref{eq:inequalities}).
The second inequality can similarly be proven by noting that for any
solution of Problem 2b, there is at least one subset $a_{i_1},
\ldots,a_{i_{n-1}}$ of $n-1$ elementary transpositions which generate
the full group.  Since there are $(2n-2)!/(n-1)!^2$ possible such
subsets, and for each subset one can consider the same decomposition
as for the lower bound, the number of solutions has an upper bound
given by the second inequality in (\ref{eq:inequalities}).

\section{Free Energies of String Theories}
\setcounter{equation}{0}
\baselineskip 18.5pt

We will now  analyze the convergence properties of the free energies
of a variety of the string theories defined in Section 2.  The
quantity we will be interested in is the leading term in the free
energy on the sphere,
\begin{equation}
F (A) = \lim_{N \rightarrow \infty} \frac{\ln Z (0,A,N)}{N^2} .
\end{equation}
The full partition function of the string theory includes maps with a
disconnected world sheet.  By taking the logarithm, we get a sum over
all maps with a connected world sheet, as is standard in quantum field
theories.  The leading term in the free energy is of order $N^2$, and
is given by a sum over maps from $S^2 \rightarrow S^2$; it is this
term $F (A)$ which we expect will show the interesting phase structure
which we will study by analyzing the convergence properties of the
string sum.  We will now proceed to calculate this function explicitly
as a string sum for a variety of choices of types of allowed
singularities.

\vspace{0.2in}

\noindent {\mbox{\boldmath  $F_{O1}$}}

\begin{figure}
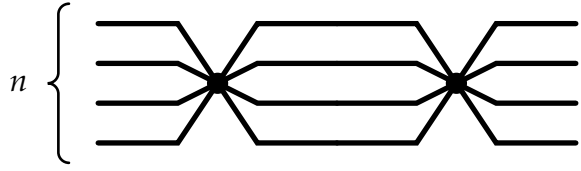

\vspace{100pt}

\caption[x]{\footnotesize String of degree $n$ contributing to $F_{O1}$}
\label{f:o1}
\end{figure}

We begin by considering the simplest possible nontrivial theory, in
which we only permit orientation-preserving maps (1 sector), and we
only allow $\Omega$-point type singularities.  This extremely simple
theory is closely related to the topological field theory used as a
starting point in \cite{cmr} to derive a world sheet formulation for
QCD${}_2$.  For a fixed degree $n$, we can easily count the number of
string maps.  Assuming that the 2 $\Omega$-points are at the poles of
the sphere, we can describe each map by the permutation on sheets
associated with the equator.  This permutation must be precisely the
permutation given at each $\Omega$-point, and furthermore must be a
cyclic permutation on $n$ sheets, or the string world sheet would be
disconnected.  Thus, for each $n$ there is a single string map, with a
symmetry factor of $n$ corresponding to the cyclic rotation of the
sheets (a schematic picture of such a map is given in
Figure~\ref{f:o1}; in the schematic representations of string maps we
will always assume that the ends of the sheets on the left and right
are connected).  We have then an exact expression for the free energy
\begin{eqnarray}
\energy{O1}(A)  & = & \sum_{n =1}^{\infty}  \frac{1}{n}
{\rm e}^{-\frac{nA}{2} }
\nonumber\\
 & = &  -\ln (1-{\rm e}^{-\frac{A}{2} }).
\end{eqnarray}
This free energy is finite and smooth away from $A = 0$; thus, a
theory with 1 sector in which only $\Omega$-point singularities occur
has no phase transitions.

It is amusing to note that the free energy of this theory
is the same as that of the trivial (complex) matrix model with
a partition function given
by
\begin{equation}
Z = \frac{1}{Z_0} \int {\cal D} M {\rm e}^{-{\rm Tr}\; (MM^{\dagger})
+ g {\rm Tr}\; (MM^{\dagger})},
\end{equation}
where $g = \exp (-A/2)$.

\vspace{0.2in}

\noindent {\mbox{\boldmath  $F_{O2}$}}

\begin{figure}
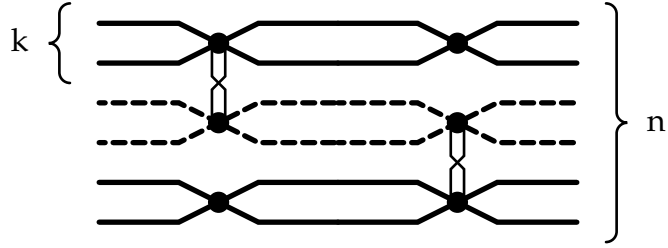

\vspace{130pt}

\caption[x]{\footnotesize String of degree $n$ contributing to $F_{O2}$}
\label{f:o2}
\end{figure}

We now consider the theory with only $\Omega$-point singularities but
with both sectors of maps allowed.  Now, we find that for fixed $n$,
there are 2 distinct string maps for each $k | n$.  Each such map has
a symmetry factor of $1/k$ and carries a sign $ (-1)^{n/k-1}$.  An
example of such a map is shown schematically in Figure~\ref{f:o2}.
The free energy for this theory is then given by
\begin{eqnarray}
\energy{O2}(A)  & = & -2 \sum_{n =1}^{\infty} \sum_{k | n}
(-1)^{n/k}\frac{1}{k}  {\rm e}^{-\frac{nA}{2} }  \nonumber\\
 & = &  2 \sum_{m = 1}^{\infty}(-1)^m  \ln (1-{\rm e}^{-\frac{mA}{2} }).
\end{eqnarray}
Again, the free energy  converges to a finite value and is smooth away
from $A = 0$, so there is only a single phase to this string theory.
It seems, then, that $\Omega$-point singularities alone are not
responsible for a phase transition in the full theory.
This result was also derived by Douglas in \cite{douglas} using a
description of QCD${}_2$ in terms of bosonic fields.

\vspace{0.2in}

\noindent {\mbox{\boldmath  $F_{Bi}$}}

Now let us drop the $\Omega$-point singularities and consider a theory
with only branch-points,  and a single sector.  For each value of $n$,
the number of maps of degree $n$ is precisely $G_n$.  Thus, using
the exact answer to Problem 2, we have
\begin{eqnarray}
\energy{B1}(A)  & = &  \sum_{n =1}^{\infty}    \frac{G_n}{(2n-2)!}  A^{2n-2}
 {\rm e}^{-\frac{nA}{2} }  \nonumber\\
 & = &  \sum_{n =1}^{\infty}    \frac{n^{n-3}}{n!}  A^{2n-2}
 {\rm e}^{-\frac{nA}{2} }.\label{eq:freebranch}
\end{eqnarray}
To study the convergence properties of this free energy, we use the
Stirling approximation for large $n$
\begin{equation}
n!  \sim \sqrt{2 \pi n}\; n^n {\rm e}^{-n}.
\end{equation}
The free energy will thus behave for large $n$ like
the series
\begin{equation}
\energy{B1} (A) \sim \sum_{n = 1}^{\infty}  \frac{[g (A)]^n }{n^{7/2}},
\end{equation}
where
\begin{equation}
g (A) = A^2 \exp (1-A/2).
\end{equation}
This series is convergent when $g (A)\leq 1$, which holds for $A \leq
A_0$ and $A \geq A_1$ with $A_0 \sim 0.73$, $A_1 \sim 11.9$.
Thus, we have a free energy which diverges for
\begin{equation}
A_0 < A < A_1
\end{equation}
and converges outside this range.  Note that although we have used the
exact result
(\ref{eq:solution2}), the bounds (\ref{eq:inequalities}) are
sufficient to prove this result qualitatively; these bounds
translate into upper and lower bounds on the  points $A_0,A_1$ such
that the upper bound on $A_0$ is much less than the lower bound on
$A_1$.

Thus, we find that simply including branch-point singularities into
the string theory is sufficient to drive a divergence of the string
sum.  Moreover, the theory with only branch-point singularities seems
to divide into 3 distinct regions, with the large $A$ and small $A$
regions {\it both} being described by a convergent string expansion.

One might also wish to consider the theory with branch-points and two
sectors.  However, because branch-point singularities by themselves
cannot connect sheets of opposite orientation, we have the rather
trivial result that
\begin{equation}
\energy{B2}(A)= 2\energy{B1}(A).
\end{equation}
Thus, adding in both chiral sectors does not have a significant effect
on the free energy of this theory.

\vspace{0.2in}

\noindent {\bf Tubes}

\begin{figure}
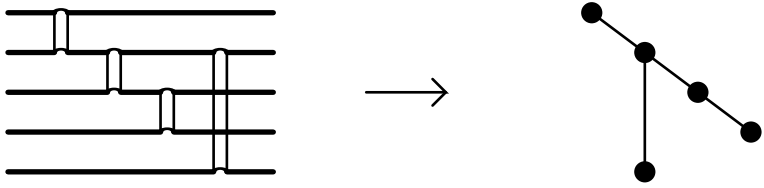

\vspace{100pt}

\caption[x]{\footnotesize Graph corresponding to string with tubes}
\label{f:tubes1}
\end{figure}

Let us now consider the effects of allowing infinitesimal tube
singularities.  First, consider the theory $F_{T1}$ with only a single
sector and orientation-preserving tubes.  For a degree $n$ map to have
a genus 0 world sheet, there must be precisely $n-1$ tubes, connecting
the sheets in a configuration corresponding to a connected graph (see
Figure~\ref{f:tubes1}).  Thus, for a fixed set of $n-1$ tube
singularity positions on the target space (corresponding to a labeling
of the edges of the associated graph), there are precisely $C_n/n=
n^{n-3}$ distinct string maps.
Since the positions of the tubes are
distinct for all but a set of configurations of measure 0, the symmetry
factor is 1 for all maps, and the associated free energy is given by
\begin{equation}
\energy{T1}(A)=\sum_{n =1}^{\infty}    \frac{n^{n- 3}}{ (n-1)!}
  A^{n-1} {\rm e}^{-\frac{nA}{2} }.
\end{equation}
If we include both sectors, the number of maps simply doubles, and we
have
\begin{equation}
\energy{T2}(A)=2 \energy{T1}(A)
\label{eq:freetubes}
\end{equation}
As $n \rightarrow \infty$, these expansions behave in an analogous
fashion to the free energy $\energy{B1}$ in (\ref{eq:freebranch}).
Here, however, the expansion diverges when
\begin{equation}
A \exp (1-A/2) > 1,
\label{eq:t}
\end{equation}
which occurs in the approximate range
\begin{equation}
0.46 < A < 5.36
\end{equation}
Thus, this theory is also well-defined for small and large $A$ but
diverges for intermediate areas.

Now, let us consider the effects of orientation-reversing tubes.  Of
course, we need both sectors of the theory for  this type of
singularity to exist, so the simplest theory containing these objects
is the theory $F_{\tilde{T}2}$.  The set of maps with only
orientation-reversing tube singularities is in a 1-1 correspondence
with maps having orientation-preserving tubes; however, now each tube
carries a factor of $-1$, so the free energy is given by
\begin{equation}
\energy{\tilde{T}2}(A)=2 \sum_{n =1}^{\infty} \frac{n^{n- 3}}{ (n-1)!}
  (-A)^{n-1} {\rm e}^{-\frac{nA}{2} }.
\end{equation}
This series has a radius of convergence identical to that of
(\ref{eq:freetubes}); however, note that now the consecutive terms in
the series alternate in sign.

\begin{figure}
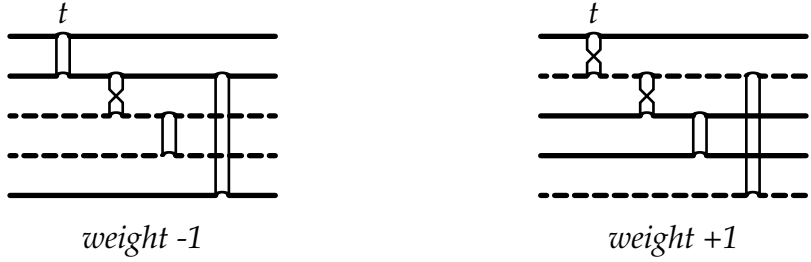

\vspace{132pt}

\caption[x]{\footnotesize Canceling strings with opposite orientation
tubes $t$}
\label{f:cancel}
\end{figure}

We now consider a theory in which both orientation-preserving and
orientation-reversing tubes are present.  In the simplest such theory,
where no other singularities exist, it is easy to see that the 1-1
correspondence between the sets of maps containing only tubes of  a
single type gives rise to a cancellation between all maps with tubes.
Specifically, if we have a map of degree $n$, with a particular tube
$t$ which is orientation-preserving, there exists a complementary
map with an orientation-reversing tube at $t$, and with all sheets
on one side of the tube reversed in orientation.  These maps have
opposite weights and therefore cancel.   An example of
such a canceling pair of maps is shown in Figure~\ref{f:cancel}.
As a
result of this cancellation, we have
\begin{equation}
\energy{T \tilde{T}2} = 2{\rm e}^{-A/2},
\end{equation}
with the only contribution arising from the two maps without tubes.

It is important to note that the cancellation described here does not
hold to all orders of the $1/N$ expansion.  If we consider terms
corresponding to a toroidal world sheet, for instance, there are
contributions from maps where the set of tubes form a closed loop on
the set of world sheets.  In this case, the number of tubes in the
loop must be even, and a single tube cannot be reversed in orientation
since the relative orientation of the sheets it is connected to is
determined independently by the remaining tubes in the loop.  Thus,
the cancellation of tubes is only an effect at highest order.

What if we allow other singularity types as well as the two types of
tubes in our theory?  In fact, it is fairly easy to see that the
cancellation described here still holds regardless of the other
singularity types allowed, since all sheets on one side of a tube can
be reversed without affecting the weight of the string map.  Thus, we
find that in any theory with both orientation-preserving and
orientation-reversing tubes, the tubes have absolutely no effect on
the leading order free energy term $F(A)$.  A particularly interesting
example of this calculation relates the full $SU(N)$ free energy to
the $Q = 0$ sector of the full $U(N)$ theory,
\begin{equation}
F_{SU(N)}= F_{BHOT \tilde{T}2}=F_{BO2}=F_{U(N)^{Q = 0}}.
\label{eq:equivalence}
\end{equation}
This important result indicates that the DK analysis of the $Q = 0$
sector of the $U(N)$ theory applies directly to the complete $SU(N)$
theory. An argument for this equivalence was also given by Minahan and
Polychronakos in the matrix model formulation \cite{mp}; however,
their argument, which states that the $n^2/N$ term in the quadratic
Casimir is negligible for large $N$, does not deal carefully with the
fact that when $n\sim N^2$, which is true at the saddle point, this
term is of the same order as the term giving rise to branch-points.
One may consider the proof given here as a precise restatement of
their argument in the string language.  This result should not be too
surprising, since we expect the $U(1)$ effects in $SU(N)$ to be of
order $1/N$ relative to the $U(N)$ effects; nonetheless, it is
gratifying to have such a simple proof of this equivalence in the
string language.

\vspace{0.2in}

\noindent {\bf Chiral theory}

We will now turn our attention to the more complicated chiral theory
with free energy $F_{BO1}$.  This theory contains all the singularity
types of the $U(N)$ theory, and is of particular interest because the
full $U(N)$ theory essentially factorizes into two copies of this
theory, connected only by the orientation-reversing tubes at
$\Omega$-points.  An analytic calculation of all terms in the free
energy of this theory using techniques such as those above is quite
difficult, so we use numerical methods here to study the convergence
properties of this series.  An analytic expression for the free energy
of this theory can be calculated using a matrix model, and will be
given in \cite{ct}.

We can write the free energy for the chiral theory in the form
\begin{equation}
F_{BO1} (A) = \sum_{n = 1}^{\infty}   \phi_n (A) {\rm e}^{-\frac{nA}{2} },
\end{equation}
where the functions $\phi_n (A)$ are polynomials of degree $2n-2$ in
$A$.  For the first few values of $n$, these polynomials are given by
\begin{eqnarray}
\phi_2 (A) & = &\frac{1}{2} - A + \frac{1}{4} A^2 \nonumber \\
\phi_3 (A) & = &\frac{1}{3} - 2 A + 3 A^2 - \frac{4}{3} A^3 +
\frac{1}{6} A^4 \nonumber \\
\phi_4 (A) & = &\frac{1}{4} - 3 A + \frac{21}{2} A^2 - \frac{43}{3} A^3 +
\frac{33}{4} A^4 - 2 A^5 + \frac{1}{6} A^6 \nonumber \\
\phi_5 (A) & = &\frac{1}{5} - 4 A + 25 A^2 - \frac{202}{3} A^3 +
\frac{529}{6} A^4 -
\frac{883}{15} A^5  + \frac{121}{6} A^6
- \frac{10}{3} A^7 + \frac{5}{24} A^8  \\
\phi_6 (A) & = &\frac{1}{6} - 5 A + \frac{195}{4} A^2 -
\frac{647}{3} A^3 + \frac{1489}{3} A^4
- \frac{3178}{5} A^5 + \frac{1871}{4} A^6 - \frac{598}{3} A^7
\nonumber\\
& & +48 A^8 - 6 A^9 + \frac{3}{10} A^{10} \nonumber \\
\phi_7 (A) & = &\frac{1}{7} - 6 A + 84 A^2 - \frac{1652}{3} A^3 + 1953 A^4 -
\frac{20228}{5} A^5  + \frac{460013}{90} A^6 - \frac{141083}{35} A^7
\nonumber \\
& &+ \frac{47983}{24} A^8 - \frac{22211}{36} A^9+ \frac{4557}{40} A^{10}
- \frac{343}{30} A^{11} + \frac{343}{720} A^{12} \nonumber
\end{eqnarray}
Note that the leading coefficient in $\phi_n$ is $n^{n-3}/n!$,  and
the constant term is $1/n$,  since these terms correspond to maps
containing only branch-point and $\Omega$-point singularities,
respectively.

\begin{figure}
\epsfbox{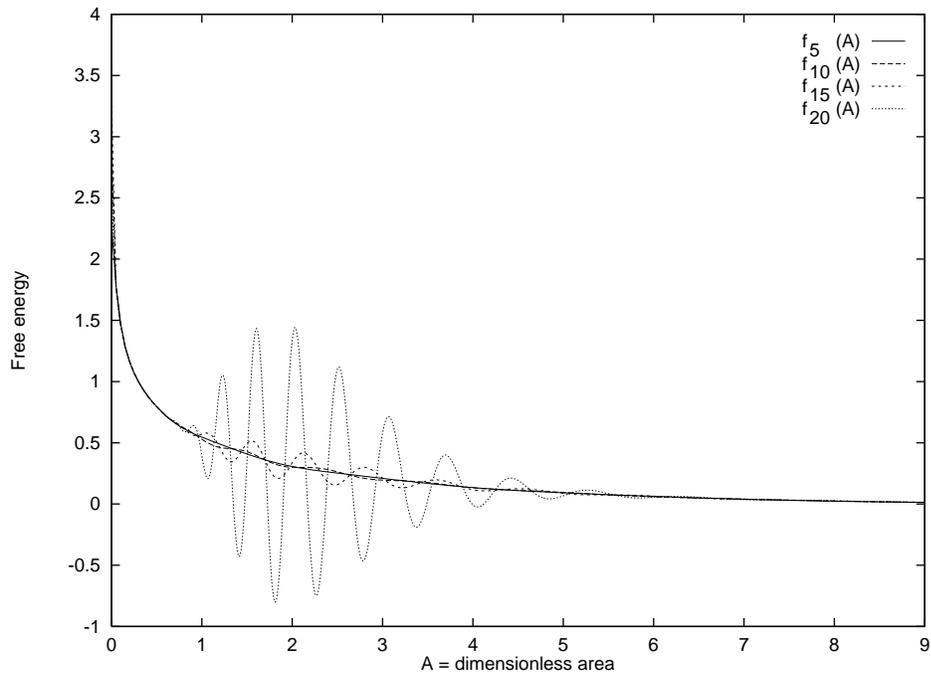}
\caption[x]{\footnotesize Partial sums of chiral free energy}
\label{f:partial1}
\end{figure}

We can write the partial summations of the string series for the free
energy as
\begin{equation}
f_n (A) = \sum_{m = 1}^{n}   \phi_m (A) {\rm e}^{-\frac{mA}{2} }
\end{equation}
To investigate the convergence properties of the string series, we  have
calculated the polynomials $\phi_n$ for $n \leq 20$, and looked at the
convergence properties of the sequence $f_n (A)$.  In
Figure~\ref{f:partial1}, the partial string summations are graphed for
$n =  5,10,  15,  20$.  It is clear from this graph that
just as for the theory without $\Omega$-points, the string summation
seems to converge for large $A$ and small $A$,  and to have  an
increasing amplitude of oscillation for intermediate areas.  A more
careful examination of the functions $f_n$ in the small and large area
regions indicates that the oscillations increase in amplitude  for
areas larger than approximately $0.5$ and below approximately $10$.
The oscillatory behavior of these partial summations is precisely
analogous to that which occurred in the theory with only
orientation-reversing tubes, and seems to indicate that in this chiral
$ U(N) $ theory, there are again 3 distinct regions, with the small
and large area regions being described by a convergent string
summation.

\vspace{0.2in}

\noindent {\bf Full SU(N) theory}

\begin{figure}
\epsfbox{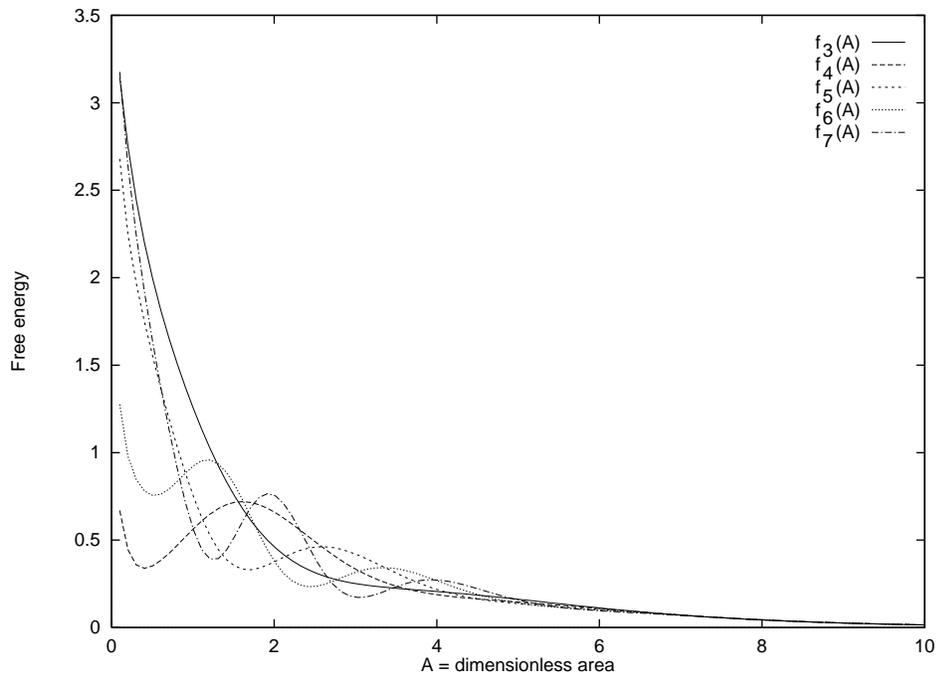}
\caption[x]{\footnotesize Partial sums of full $SU(N)$ free energy}
\label{f:full}
\end{figure}

We can use the same technique of computing partial summations of the
string series to analyze the free energy $F_{BO2}$, which by
(\ref{eq:equivalence}) is the complete leading term in the free energy
of the $SU(N)$ theory, and which is precisely the theory analyzed by
DK using matrix models.  Graphing the first few partial sums, we have
Figure~\ref{f:full}.

{}From this figure, we might conclude that the string expansion is
only valid in the large $A$ phase, and that the partial summations
have oscillations of increasing amplitude for small areas.  Further
numerical analysis of the partial summations indicates that the series
ceases to converge around $A = 10$.  It is difficult to show that this
occurs precisely at the DK phase transition point; however, this
result indicates that the strong coupling string expansion may contain
sufficient information to determine the phase transition point at $A =
\pi^2$.  For this theory only a small number of terms need to be
considered to see the qualitative behavior of two phases.  The
observation that the convergence radius of the string expansion
is related to the phase transition point of the QCD${}_2$ theory
leads us to conjecture that a similar relationship holds for the
chiral theory described above.  In this theory we would therefore
expect to find a phase
transition near $A = 10$, and
possibly also to find a third phase of the theory for small $A$.  Recent
work using a matrix model for chiral QCD${}_2$ indicates that
these conclusions are correct \cite{ct}.

Assuming that the radius of convergence of the string expansion is
indeed given by the phase transition point, it is now possible to
characterize the phase transitions in this theory as being due to the
same mechanism that causes the divergence of the string expansion in
the theory $F_{B1}$.  This divergence is driven by the competing
factors associated with the entropy of branch-points, which
asymptotically looks like $A {\rm e}^{1/2}$ for each branch-point, and
the action, which gives a factor of ${\rm e}^{-A/4}$ for each
branch-point.  When the area is sufficiently large, the factor
associated with the action becomes sufficiently small to damp the
large branch-point entropy, and when the area is very small, the
entropy factor becomes small while the action term approaches unity,
making the string expansion convergent in both these regimes.
However, in the intermediate regime, the the sum diverges because the
entropy of the branch-points overwhelms the suppression due to the
action factor.  It is tempting to conclude that this
mechanism is essentially responsible for the phase transition found in
the full 2D QCD theory.

\section{Conclusions}
\setcounter{equation}{0}
\baselineskip 18.5pt

In this paper we have examined a variety of simple string theories
which include the string descriptions of the 2D $U(N)$ and $SU(N)$
gauge theories.  For the simpler string theories containing only
$\Omega$-point or branch-point types of singularities, we have given
an analytic expression for the leading term in the free energy, and we
have found that while $\Omega$-points alone do not affect the
convergence of the string series, the entropy of branch-points can be
sufficient to cause a divergence of the string free energy.  We have
proven that the leading term in the free energy for the $SU(N)$ theory
is identical to that of the $Q = 0$ sector of the $U(N)$ theory
studied by Douglas and Kazakov using a matrix model formulation and
more recently analyzed by Gross and Matytsin in the small area phase.
We have numerically performed partial summations of the string series
for the full and chiral $U(N)$ free energy.  These results indicate
that for the full theory the string expansion converges in the large
area phase and fails to converge in the small area phase.  For the
chiral theory, the string expansion also converges for very small
areas, indicating the possible existence of 3 phases in that theory.

The apparent coincidence between the radius of convergence of the
strong coupling string expansion and the phase transition point is in
some sense a surprising result.  Because a phase transition is
essentially a non-analytic event, we do not expect to be able to
predict a transition point from a perturbative expansion around a
point in one phase.  However, the string expansions are not simple
perturbative expansions; they are double expansions in the quantities
$A$ and ${\rm e}^{-A}$ which contain some nonperturbative information.
In fact, it seems that these expansions actually do carry information
about the phase transition points.  At this time, however, it is not
clear how this association might be proven; only the numerical
evidence given here indicates the correspondence of the phase
transition point with the string radius of convergence.  This is
certainly an interesting question deserving further study.

Perhaps the most interesting of the results given here is the
prediction of two distinct phase transition points for the chiral
theory.  The chiral theory itself is quite an interesting and rather
mysterious object.  It arises naturally when we look at the large $N$
$SU(N)$ or $U(N)$ theory, where the string partition function
factorizes almost exactly into two copies of the chiral theory, with
small corrections.  The chiral theory also appears naturally in the
formalism of CMR as a topological field theory augmented with a Nambu
action \cite{cmr}.  It is furthermore possible to use the chiral
theory to describe a string theory for finite $N$ QCD by the insertion
of a simple projection operator \cite{bt}.  However, as yet we do not
really have an intrinsic definition of the chiral theory as a field
theory on the target space; it should be possible to find a
factorization of a large $N$ gauge theory into two copies of a target
space version of the chiral theory, but so far such a decomposition
has proven elusive.  A better understand of this chiral theory may
eventually be of assistance in formulating and understanding a string
description of 4D QCD.  The observations made here about the phase
structure of this theory are a first step in that direction.  An
investigation of the chiral theory using a matrix model formulation
has recently been carried out by M.\ Crescimanno and the author; this
work also indicates the existence of 3 phases in the chiral theory,
and will be reported separately \cite{ct}.

This work may also contribute some relevant insights to a better
physical understanding of the QCD${}_2$ phase transition.  In the work
of Gross and Matytsin \cite{gm} it was shown that when viewed from the
small area phase of the 2D gauge theory, the DK phase transition is
driven by instantons.  The results here indicate that when one
approaches the phase transition point from the other side, that is
from the large area phase, the phase transition is driven by the
entropy of branch-point singularities, which dominates over the
Boltzmann weight of these objects.  This result, in combination with
the results of Gross and Matytsin, indicates that it may be possible
to find a direct connection between the branched coverings of the
string picture and instantons in the gauge theory picture.  Such a
connection, if made rigorous, would greatly strengthen our
understanding of string formulations of gauge theories, and is a
promising direction for future work.

It is hoped that the work in this paper will make it clear that there
are simple and important observations which can be made from the
string approach to QCD which are not as transparent using other
methods.  In order to develop a useful string formulation of QCD in 4
dimensions, it is important not only to work on a formal description
of gauge theories as strings, but also to develop the computational
tools necessary to allow us to compute phenomena from the string point
of view which might be inaccessible from a pure gauge theory
perspective.  The work here represents a modest step in that
direction.

\vskip .5truein
{\Large{\bf Acknowledgements}}

The author would like to thank M.\ Douglas, D.\ Gross, I.\ Singer, and
C.\ Waldman for helpful discussions and correspondence.  Particular
thanks to M.\ Crescimanno for suggesting several improvements in this
manuscript and for interesting and useful discussions.

\end{document}